\documentclass{article}

\usepackage{spconf}
\usepackage{amsmath}
\usepackage{amssymb}
\usepackage{amsthm}
\usepackage{graphicx}
\usepackage{bm}
\usepackage{booktabs}
\usepackage{multirow}
\usepackage{hyperref}
\usepackage{orcidlink}
\usepackage{titlesec}

\DeclareMathOperator{\tr}{tr}
\DeclareMathOperator*{\argmin}{arg\,min}

\newtheorem{theorem}{Theorem}
\newtheorem{corollary}{Corollary}

\title{Bayes-Optimal Quantum Computational Sensing \\ via Convex Spectral Design}

\name{%
Amirhossein Taherpour~\orcidlink{0000-0003-4647-102X},
Alireza Sadeghi~\orcidlink{0000-0003-1280-7592},
and Georgios B.~Giannakis~\orcidlink{0000-0002-0196-0260}
\thanks{%
The authors are with the Department of ECE,
University of Minnesota, Minneapolis, MN, USA
(e-mails: {taher054,sadeg012,georgios}@umn.edu).
This work was supported in part by NSF grants 2212318 and 2312547.
Corresponding author: Georgios B.~Giannakis.%
}
}
\address{%
Department of Electrical and Computer Engineering, University of Minnesota, USA
}

\begin{document}
\maketitle

\begin{abstract}
Quantum computational sensing performs task-specific inference directly
through coherent sensing dynamics, avoiding an intermediate estimation step.
Existing designs, however, typically optimize quantum control parameters
through nonconvex programs whose performance can depend strongly on
initialization. This contribution formulates binary phase classification directly in terms of
the sensor's measurement response. For a fixed query budget, generalized
quantum signal processing (GQSP) restricts this response to a bounded trigonometric
polynomial, while the Bayes classification error is linear in the response.
These properties yield an exact finite-dimensional semidefinite program for
the globally Bayes-optimal sensor response. Fej\'er--Riesz factorization and
GQSP synthesis then recover a physically realizable control sequence. The
same formulation extends naturally to sample-based design and additive phase
uncertainty. Numerical tests show improved accuracy and optimization
reliability over direct control-angle training; noise-aware redesign also
lowers error under phase uncertainty.
\end{abstract}


\section{Introduction}
\label{sec:intro}

Quantum sensing uses controlled quantum dynamics to infer physical signals
with high sensitivity
\cite{Degen2017QuantumSensing,Vouras2023QuantumSyntheticApertures,
Gong2023QuantumTransmittance}. Recent scalable
quantum-state reconstruction includes distributed tomography robust to readout
errors and rank-adaptive matrix-free low-rank tomography
\cite{Taherpour2026DistributedQST,Taherpour2026AtomicQST,
Shanmugam2023LQST,Xue2025HybridQST}. Many applications,
however, seek a task-specific decision rather than the underlying parameter or
state \cite{Lee2023QuantumBayesian,Taherpour2025RAPID}.
Quantum computational sensing (QCS) embeds task-dependent processing in
coherent sensing dynamics so the final measurement directly produces the
desired inference.

Quantum signal processing (QSP) and generalized quantum signal processing
(GQSP) realize structured polynomial responses to encoded signals
\cite{Low2017QSP,Motlagh2024GQSP,Berntson2025Complementary,
Lu2026UNQSP,Liu2025MixedAnalogDigitalQSP}.
Recent advances span infinite QSP, efficient GQSP simulation, stochastic QSP,
single-shot interferometry, low-depth phase estimation, and programmable signal
design
\cite{Dong2024InfiniteQSP,Berry2024GQSP,Martyn2025StochasticQSP,
SinananSingh2024QSPI,Dong2025QSPE,Jia2026Programmable}.
Recent QCS theory and experiments further cover binary, multiclass, and
time-dependent sensing on qubit and hybrid platforms
\cite{Khan2026QCS,Prabhu2026Displacement,Sen2026QuantumSensors}.
Such designs typically train parameterized controls directly, yielding
nonconvex optimization that can depend strongly on initialization
\cite{Khan2026QCS,Sen2026QuantumSensors}.

We instead optimize the measurement response the quantum sequence must realize.
For binary phase classification at a prescribed query budget, GQSP restricts
this response to a bounded trigonometric polynomial, while the Bayes
classification error is linear in the response. Nonnegative
trigonometric-polynomial representations therefore yield an exact
finite-dimensional semidefinite program (SDP), replacing direct nonconvex
control optimization by globally optimal convex response-domain design.
Fej\'er--Riesz factorization recovers polynomial amplitudes
\cite{Dritschel2010FejerRiesz}, and standard GQSP synthesis maps them to a
realizable control sequence
\cite{Motlagh2024GQSP,Berntson2025Complementary}. The framework also admits
sample-based design and additive phase uncertainty without destroying
convexity. Numerical results demonstrate improved classification performance
and reduced sensitivity to control-angle initialization.

\section{Convex spectral design}
\label{sec:design}

\noindent \textbf{Task-specific quantum sensing.} Let \(Y\in\{0,1\}\) denote the random class label and \(y\) a
realization of \(Y\), with \emph{prior} probability
\(\Pr(Y=y)=\pi_y\), where \(\pi_0+\pi_1=1\).
Let \(\Theta\in[-\pi,\pi)\) denote the \emph{unknown} physical quantum-phase and
\(\theta\) a realization of \(\Theta\). Conditioned on \(Y=y\), the conditional quantum-phase
has density
\(f_{\Theta\mid Y}(\theta\mid y)=f_y(\theta)\).
The interval \([-\pi,\pi)\) represents one \(2\pi\)-period of the phase.
Rather than first estimating \(\Theta\) and subsequently classifying it,
we seek a \emph{single-shot} quantum sensor whose binary measurement
\(\widehat Y\in\{0,1\}\) directly predicts \(Y\). Under the standard single-qubit GQSP architecture
\cite{Motlagh2024GQSP}, the unknown phase enters through the
signal-dependent operation
\(\mathbf S(\theta)=\operatorname{diag}(e^{\mathrm{j}\theta},1)\),
where \(\mathrm{j}\triangleq\sqrt{-1}\). A \(Q\)-query sequence contains
\(Q\) applications of \(\mathbf S(\theta)\), interleaved with
\(Q+1\) signal-independent rotations
\(\mathbf R_q\in\mathrm{SU}(2)\), \(q=0,\ldots,Q\), where
\(\mathrm{SU}(2)\) denotes the \(2\times2\) special unitary group.
Starting from \(\lvert0\rangle\) and measuring in the computational basis yields
\begin{equation}
\begin{aligned}
\mathbf U_Q(\theta)
=& \;
\mathbf R_Q\mathbf S(\theta)\mathbf R_{Q-1}
\cdots
\mathbf S(\theta)\mathbf R_0,
\\
p(\theta)
\triangleq &\;
\Pr(\widehat Y=1\mid\Theta=\theta)
=
\left|
\langle1|\mathbf U_Q(\theta)|0\rangle
\right|^2 .
\end{aligned}
\label{eq:qcs}
\end{equation}
Thus, \(p(\theta)\) is the sensor's class-\(1\) response, while
\(\Pr(\widehat Y=0\mid\Theta=\theta)=1-p(\theta)\). The GQSP polynomial characterization
implies that the
\(Q\) occurrences of \(\mathbf S(\theta)\) in \eqref{eq:qcs} restrict
the response to a real trigonometric polynomial of degree at most \(Q\) \cite{Motlagh2024GQSP,Berntson2025Complementary}.
Let this trigonometric-response space be
\[
\mathcal T_Q
\triangleq
\Bigg\{
p(\theta)
=
\sum_{k=-Q}^{Q}
c_k e^{\mathrm{j}k\theta}:
c_{-k}=c_k^*
\Bigg\}
\]
where \(k\) is the harmonic index and \(c_k\) is the corresponding Fourier coefficient. The conjugate symmetry \(c_{-k}=c_k^*\) makes
\(p\) real valued. Moreover, because \(\mathbf U_Q(\theta)\) is
unitary, the Born-rule probability in \eqref{eq:qcs} satisfies
\(0\le p(\theta)\le1\) for every \(\theta\). We show next that these
conditions are also sufficient for GQSP realizability. We next select the
realizable response that minimizes classification error.

For a response \(p\), define the classification error
\(\mathcal E(p)\triangleq\Pr(\widehat Y\neq Y)\).
An error occurs either when \(Y=0\) and the sensor reports \(1\), or when
\(Y=1\) and it reports \(0\). Therefore,
\[
\mathcal E(p) =
\pi_0
\int_{-\pi}^{\pi}
f_0(\theta)p(\theta)\,d\theta
+
\pi_1
\int_{-\pi}^{\pi}
f_1(\theta)\bigl[1-p(\theta)\bigr]\,d\theta
\]
thus $\mathcal E(p) = \pi_1+
\int_{-\pi}^{\pi}
w(\theta)p(\theta)\,d\theta$,
where
\(w(\theta)\triangleq
\pi_0f_0(\theta)-\pi_1f_1(\theta)\)
is the signed Bayes weight. Its sign has a direct classification
interpretation: without the finite-\(Q\) restriction, Bayes rule
selects class \(1\) for \(w(\theta)<0\), and \(0\) for \(w(\theta)>0\).

The Bayes-optimal response achievable with \(Q\) signal queries therefore
solves
\begin{equation}
\begin{aligned}
p_Q^\star
\in
\argmin_{p\in\mathcal T_Q}
\quad&
\left[ \pi_1+
\int_{-\pi}^{\pi}
w(\theta)p(\theta)\,d\theta \right]
\\
\operatorname{subject\ to} \quad&
0\le p(\theta)\le1,
\qquad
\forall\theta\in[-\pi,\pi).
\end{aligned}
\label{eq:main_problem}
\end{equation}
Although \(p\) has only finitely many Fourier coefficients, the bounds in
\eqref{eq:main_problem} must hold for every
\(\theta\in[-\pi,\pi)\), and hence represent a continuum of inequalities.
The next subsection replaces these infinitely many constraints by an exact
finite-dimensional semidefinite representation.

\noindent \textbf{Exact semidefinite reformulation.} To encode the nonnegativity constraints in \eqref{eq:main_problem}, set
\(z=e^{\mathrm{j}\theta}\) and define the degree-\(Q\) vector
$
\mathbf m_Q(z)
\triangleq
[1,z,\ldots,z^Q]^{\mathsf T}.
$
For a Hermitian matrix
\(\mathbf G\in\mathbb C^{(Q+1)\times(Q+1)}\), the
\(k\)-th diagonal-sum functional is
\[
d_k(\mathbf G)
\triangleq
\sum_{r=0}^{Q-k}
[\mathbf G]_{r,r+k},
\qquad
k=0,\ldots,Q.
\]
Expanding the quadratic form gives
\[
\mathbf m_Q^\dagger(z)\mathbf G\mathbf m_Q(z)
=
d_0(\mathbf G)
+
2\operatorname{Re}
\left\{
\sum_{k=1}^{Q}
d_k(\mathbf G)z^k
\right\}.
\]
Thus, \(d_k(\mathbf G)\) determines the \(k\)-th harmonic of the
associated trigonometric polynomial. In particular, if
\(\mathbf G\succeq\mathbf0\), then
\(\mathbf m_Q^\dagger(z)\mathbf G\mathbf m_Q(z)\ge0\) for every
\(|z|=1\). Positive semidefinite Gram matrices can therefore encode
nonnegative trigonometric responses.

We use \(\mathbf G_1\) as a Gram matrix for \(p\) and
\(\mathbf G_0\) as a Gram matrix for \(1-p\), matching the two possible
measurement outcomes. To express the Bayes objective in the same
coordinates, define the task-weight matrix
\[
\mathbf W
\triangleq
\int_{-\pi}^{\pi}
w(\theta)
\mathbf m_Q(e^{\mathrm{j}\theta})
\mathbf m_Q^\dagger(e^{\mathrm{j}\theta})
\,d\theta
\]
for
\(p(\theta)=
\mathbf m_Q^\dagger(e^{\mathrm{j}\theta})
\mathbf G_1
\mathbf m_Q(e^{\mathrm{j}\theta})\),
trace cyclicity yields
\[
\int_{-\pi}^{\pi}w(\theta)p(\theta)\,d\theta
=
\tr(\mathbf W\mathbf G_1).
\]
The following result establishes that these matrix representations exactly eliminate the continuum constraints in \eqref{eq:main_problem}.

\begin{theorem}
\label{thm:exact}
The one-shot responses achievable by the \(Q\)-query GQSP sensor
\eqref{eq:qcs} are exactly
\[
\left\{
p\in\mathcal T_Q:
0\le p(\theta)\le1,\ \forall\theta
\right\}.
\]
Consequently, \eqref{eq:main_problem} is equivalent to
\begin{equation}
\begin{aligned}
\underset{\mathbf G_0,\mathbf G_1}{\operatorname{minimize}}
\quad&
\pi_1+\tr(\mathbf W\mathbf G_1)
\\
\operatorname{subject\ to}
\quad&
\mathbf G_0\succeq\mathbf0,
\qquad
\mathbf G_1\succeq\mathbf0,
\\
&
d_0(\mathbf G_0+\mathbf G_1)=1,
\\
&
d_k(\mathbf G_0+\mathbf G_1)=0,
\qquad k=1,\ldots,Q .
\end{aligned}
\label{eq:sdp}
\end{equation}
If
\((\mathbf G_0^\star,\mathbf G_1^\star)\)
solves \eqref{eq:sdp}, then
$
p_Q^\star(\theta)
=
\mathbf m_Q^\dagger(e^{\mathrm{j}\theta})
\mathbf G_1^\star
\mathbf m_Q(e^{\mathrm{j}\theta})$
is globally optimal for \eqref{eq:main_problem} and is realizable with
\(Q\) signal queries.
\end{theorem}

\begin{proof}
A \(Q\)-query GQSP sequence produces polynomial amplitudes
\(a_0(z)\) and \(a_1(z)\) of degree at most \(Q\) satisfying
\[
|a_0(z)|^2+|a_1(z)|^2=1,
\qquad |z|=1.
\]
Hence, the class-\(1\) response
\(p(\theta)=|a_1(e^{\mathrm{j}\theta})|^2\)
belongs to \(\mathcal T_Q\) and satisfies \(0\le p\le1\).

For the converse, the Fej\'er--Riesz theorem
\cite{Dritschel2010FejerRiesz} states that any nonnegative
trigonometric polynomial of degree at most \(Q\) can be expressed on the
unit circle as the squared magnitude of an ordinary polynomial of degree
at most \(Q\). Since \(0\le p\le1\), both \(p\) and \(1-p\) are nonnegative, there exist degree-\(Q\) polynomials \(a_1\) and \(a_0\)
such that
\[
p(\theta)=|a_1(e^{\mathrm{j}\theta})|^2,
\qquad
1-p(\theta)=|a_0(e^{\mathrm{j}\theta})|^2.
\]
Their sum is one on the unit circle, so \(a_0\) and \(a_1\) form a
complementary polynomial pair. Standard GQSP synthesis maps such a pair to a
\(Q\)-query sequence of rotations, establishing physical realizability
\cite{Motlagh2024GQSP,Berntson2025Complementary}. Thus, the same factorizations yield positive semidefinite Gram representations
\[
p(\theta)
=
\mathbf m_Q^\dagger\mathbf G_1\mathbf m_Q,
\qquad
1-p(\theta)
=
\mathbf m_Q^\dagger\mathbf G_0\mathbf m_Q.
\]
The identity \(p+(1-p)=1\) requires the constant harmonic of
\(\mathbf G_0+\mathbf G_1\) to equal one and every nonconstant harmonic
to vanish, which is exactly
\[
d_0(\mathbf G_0+\mathbf G_1)=1,
\qquad
d_k(\mathbf G_0+\mathbf G_1)=0,\quad k=1,\ldots,Q.
\]
Finally, the definition of \(\mathbf W\) converts the Bayes term to
\(\tr(\mathbf W\mathbf G_1)\). Thus \eqref{eq:sdp} and
\eqref{eq:main_problem} have identical feasible responses and objective
values.
\end{proof}

Theorem~\ref{thm:exact} resolves the continuum difficulty identified
after \eqref{eq:main_problem}: the infinitely many probability bounds are
replaced exactly by two positive semidefinite constraints and
\(Q+1\) linear equality constraints. Hence, \eqref{eq:sdp} is an exact
convex reformulation rather than a discretization or relaxation. The reformulation also closes the loop from response design back to the
physical sensor. Solving \eqref{eq:sdp} first yields the globally optimal
response \(p_Q^\star\); Fej\'er--Riesz factorization then produces the
complementary amplitudes \(a_0\) and \(a_1\), and standard GQSP synthesis
maps them to the rotations
\(\{\mathbf R_q\}_{q=0}^{Q}\) in \eqref{eq:qcs}. As a result, the challenging
nonconvex search over \(3(Q+1)\) rotation parameters is replaced by
convex optimization of the measurement response via \eqref{eq:sdp}, with the controls
recovered only after the optimum has been found.

\noindent \textbf{Sample-based design and phase uncertainty.} The formulation above assumes that the class statistics
\(\{\pi_y,f_y\}_{y=0}^{1}\) are known well enough to construct the Bayes
objective in \eqref{eq:main_problem}, or equivalently the matrix
\(\mathbf W\) in \eqref{eq:sdp}. In practice, however, one may instead have only
labeled phase samples. In addition, the phase applied in each signal query
may be corrupted by uncertainty. We next show that both cases modify only
the objective of \eqref{eq:sdp}; the set of physically realizable
\(Q\)-query responses remains unchanged.

First, suppose only a labeled data set \(\mathcal D\triangleq\{(\theta_n,y_n)\}_{n=1}^{N}\) is given.
Replacing the true error \(\mathcal E(p)\) by the empirical
misclassification error yields
\[
\begin{aligned}
\widehat{\mathcal E}(p)
&=
\frac{1}{N}
\sum_{n=1}^{N}
\left[
(1-y_n)p(\theta_n)
+
y_n\bigl(1-p(\theta_n)\bigr)
\right]
\\
&=
\widehat\pi_1+
\tr(\mathbf W_{\mathcal D}\mathbf G_1)
\end{aligned}
\]
where the empirical class-\(1\) prior is
\(\widehat\pi_1\triangleq
N^{-1}\sum_{n=1}^{N}y_n\), and
\[
\mathbf W_{\mathcal D}
\triangleq
\frac{1}{N}
\sum_{n=1}^{N}
(1-2y_n)
\mathbf m_Q(e^{\mathrm{j}\theta_n})
\mathbf m_Q^\dagger(e^{\mathrm{j}\theta_n}).
\]
Thus, the globally optimal empirical sensor is obtained from the same SDP
\eqref{eq:sdp} by replacing
\((\pi_1,\mathbf W)\) with
\((\widehat\pi_1,\mathbf W_{\mathcal D})\).

Next, suppose a common additive phase perturbation \(\nu\), independent of
the true phase \(\theta\) and with a known distribution, shifts each signal
interaction in \eqref{eq:qcs} from \(\theta\) to \(\theta+\nu\), where
addition is understood modulo \(2\pi\). Let
\(\varphi_\nu(k)\triangleq
\mathbb E[e^{\mathrm{j}k\nu}]\)
denote its characteristic function. For the Fourier representation
\(p(\theta)=\sum_{k=-Q}^{Q}c_ke^{\mathrm{j}k\theta}\), the relevant
noise-averaged response is
\begin{equation}
\begin{aligned}
p_\nu(\theta)
&\triangleq
\mathbb E_\nu[p(\theta+\nu)]=
\sum_{k=-Q}^{Q}
c_k\varphi_\nu(k)e^{\mathrm{j}k\theta}.
\end{aligned}
\label{eq:noise_response}
\end{equation}
Thus, additive phase uncertainty does not introduce new harmonics; it
multiplies the \(k\)-th Fourier coefficient \(c_k\) by
\(\varphi_\nu(k)\).

\begin{corollary}
\label{cor:noise}
Define the noise-aware task-weight matrix \(\mathbf W_\nu\)
\[
[\mathbf W_\nu]_{r,s}
=
[\mathbf W]_{r,s}\,
\varphi_\nu(r-s),
\qquad
r,s=0,\ldots,Q,
\]
where \(r\) and \(s\) denote matrix row and column indices. The globally
optimal \(Q\)-query sensor under additive phase uncertainty is obtained
from \eqref{eq:sdp} by replacing \(\mathbf W\) with
\(\mathbf W_\nu\).
\end{corollary}

\noindent For Gaussian phase uncertainty
\(\nu\sim\mathcal N(0,\sigma^2)\),
\(\varphi_\nu(k)=e^{-k^2\sigma^2/2}\), so higher-order harmonics are
attenuated more strongly. Increasing \(Q\) makes higher harmonics
available to \(p(\theta)\), allowing sharper transitions near the
unconstrained Bayes decision boundaries \(w(\theta)=0\); phase uncertainty
suppresses precisely these high-order components. The noise-aware SDP
accounts for this tradeoff directly through \(\mathbf W_\nu\).

\section{Simulations and numerical results}
\label{sec:numerical}

\textbf{Setup.} Consider two binary phase-classification tasks on
\(\theta\in[-\pi,\pi)\), with the simple task using equal priors and
\[
f_0(\theta)=\mathrm{VM}(-0.90,5.5)
\qquad
f_1(\theta)=\mathrm{VM}(0.90,5.5)
\]
where \(\mathrm{VM}(\mu,\kappa)\) denotes the von Mises density with mean
direction \(\mu\) and concentration \(\kappa\). The interleaved task uses
\(\pi_0=0.48\), \(\pi_1=0.52\), and
\[
\begin{aligned}
f_0(\theta)
&=
0.58\,\mathrm{VM}(-2.30,9.5)
+
0.42\,\mathrm{VM}(-0.20,6.5)
\\
f_1(\theta)
&=
0.46\,\mathrm{VM}(-1.18,7.0)
+
0.54\,\mathrm{VM}(1.12,10.5)
\end{aligned}
\]
whose weighted densities alternate in dominance and therefore induce
multiple Bayes decision boundaries \(w(\theta)=0\). We sweep \(Q\in\{2,4,\ldots,20\}\) and evaluate phase integrals on
\(2^{15}\) uniformly spaced points. Our advocated method solves
\eqref{eq:sdp} and synthesizes the corresponding GQSP rotations as in
\cite{Motlagh2024GQSP,Berntson2025Complementary}. We compare with:
(i) the unconstrained Bayes oracle;
(ii) Angle-GQSP, which optimizes the \(3(Q+1)\) rotation parameters using
Adam for \(2,500\) iterations from \(64\) random initializations; and
(iii) a Fej\'er-smoothed Bayes decision. For the robustness study,
\(\nu\sim\mathcal N(0,\sigma^2)\) with
\(\sigma\in[0,0.8]\) rad, and the noise-aware design uses
\(\mathbf W_\nu\) from Corollary~\ref{cor:noise}.

\begin{figure}[t]
\centering
\includegraphics[width=0.9\columnwidth]{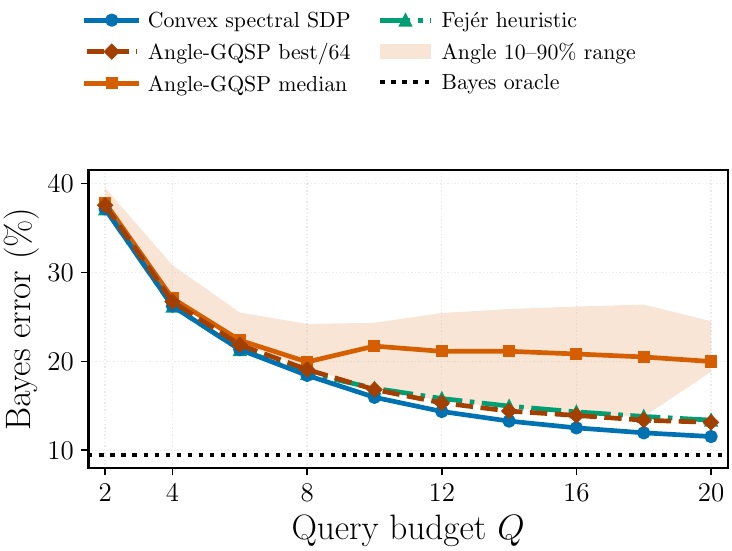}
\caption{Classification error vs. query budget (Q); Angle-GQSP shows best, median, and 10--90 percentile performance over 64 runs.}
\label{fig:depth_performance}
\end{figure}

\noindent \textbf{Query budget and optimization landscape.} Fig.~\ref{fig:depth_performance} shows that increasing \(Q\) improves classification performance by enlarging the feasible response space \(\mathcal T_Q\). Angle-GQSP has access to the same sensing architecture, but its
increasing variability across random initializations shows that favorable
rotation configurations become harder to locate as the query budget grows.
\begin{figure}[t]
\centering
\includegraphics[width=0.9\columnwidth]{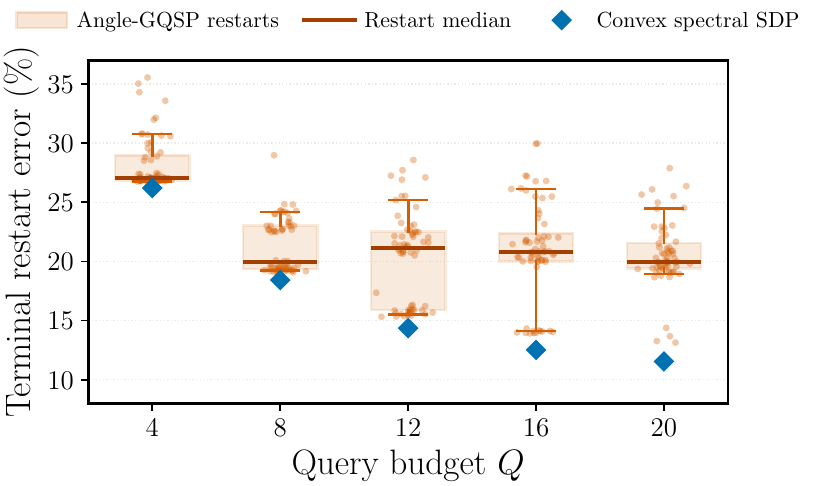}
\caption{Terminal Angle-GQSP classification errors over 64 random initializations; diamonds indicate the global SDP optimum via \eqref{eq:sdp}.}
\label{fig:restart_landscape}
\end{figure}
Fig.~\ref{fig:restart_landscape} separates this optimization effect from
representational capacity: low-error angle solutions remain attainable, but
local optimization finds them less reliably at larger \(Q\). In contrast,
the response-domain SDP optimizes globally over the feasible set characterized
in Theorem~\ref{thm:exact}.

\begin{table}[t]
\centering
\caption{Classification error (\%) and offline design time. Angle-GQSP
reports the best of \(64\) initializations. Lower is better.}
\label{tab:main_results}
\begingroup
\footnotesize
\setlength{\tabcolsep}{3pt}
\begin{tabular}{lccccc}
\toprule
&
\multicolumn{2}{c}{Simple}
&
\multicolumn{2}{c}{Interleaved}
&
\multirow{2}{*}{Time (s) $Q=16$}
\\
\cmidrule(lr){2-3}
\cmidrule(lr){4-5}
Method
& \(Q=8\) & \(Q=16\)
& \(Q=8\) & \(Q=16\)
& \\
\midrule

Bayes oracle
& 2.38 & 2.38
& 9.45 & 9.45
& -- \\

Convex SDP
& \textbf{5.56} & \textbf{3.33}
& \textbf{18.42} & \textbf{12.52}
& 3.51 \\

Angle-GQSP
& 5.94 & 3.99
& 19.07 & 13.92
& 88.37 \\

Fej\'er
& 7.97 & 5.34
& 18.65 & 14.32
& \textbf{0.012} \\

\bottomrule
\end{tabular}
\endgroup
\end{table}

\noindent Table~\ref{tab:main_results} highlights the corresponding
accuracy--design-cost tradeoff. The SDP avoids the expensive multi-start
search required by Angle-GQSP, while solving the finite-\(Q\) Bayes design
problem globally; Fej\'er smoothing is considerably cheaper but sacrifices
classification accuracy.

\begin{figure}[t]
\centering
\includegraphics[width=0.9\columnwidth]{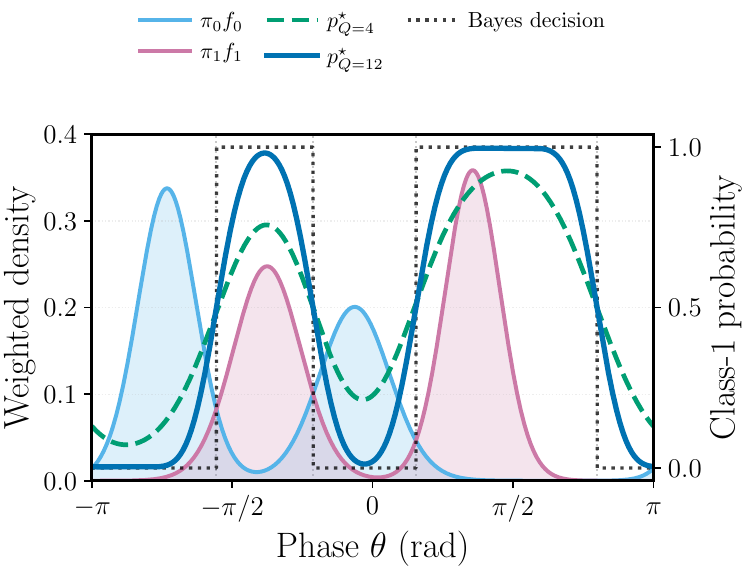}
\caption{Class-weighted densities and optimized responses; guides mark Bayes boundaries \(w(\theta)=0\).}
\label{fig:response_geometry}
\end{figure}

\noindent \textbf{Finite-query response geometry.} Fig.~\ref{fig:response_geometry} illustrates that at small \(Q\), the bandwidth constraint forces broad
transitions between neighboring class-dominant regions. Larger \(Q\)
adds higher harmonics needed to approach the sharp Bayes transitions
while maintaining \(0\le p_Q^\star(\theta)\le1\).
Across all \(Q\le20\), the synthesized GQSP response matches the SDP optimum
to within \(1.19\times10^{-8}\) pointwise, and \(2.14\times10^{-9}\) in classification error, confirming that synthesis preserves the optimized response.

\begin{figure}[t]
\centering
\includegraphics[width=0.9\columnwidth]{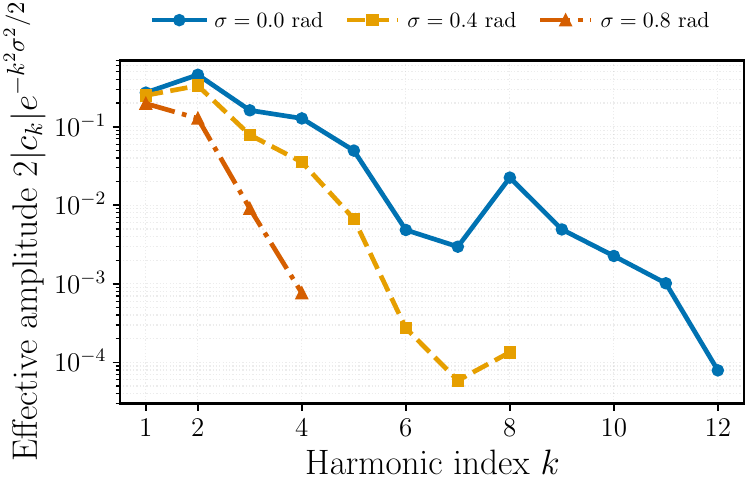}
\caption{Harmonic amplitudes for \(Q=12\) under Gaussian phase uncertainty. The \(k\)-th harmonic is attenuated by \(e^{-k^2\sigma^2/2}\).}
\label{fig:spectral_attenuation}
\end{figure}

\noindent \textbf{Phase uncertainty.}
For Gaussian phase uncertainty, \eqref{eq:noise_response} attenuates the
\(k\)-th harmonic by \(e^{-k^2\sigma^2/2}\). Fig.~\ref{fig:spectral_attenuation} depicts this spectral attenuation. Higher-order components thus
decay first, limiting sharp transitions
near \(w(\theta)=0\).

\begin{figure}[!t]
\centering
\includegraphics[width=0.9\columnwidth]{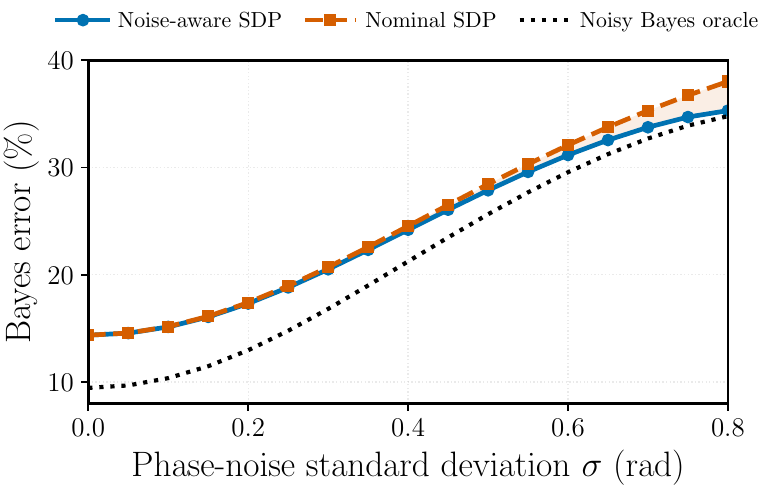}
\caption{Classification error vs. Gaussian phase noise at \(Q=12\), for noise-aware design.}
\label{fig:phase_noise_robustness}
\end{figure}

Accordingly, noise-aware redesign offers little advantage when \(\sigma\)
is small, but becomes increasingly useful as uncertainty suppresses the
high-frequency structure of the nominal response. As shown in Fig.~\ref{fig:phase_noise_robustness}, this yields increasing performance gains at larger noise levels, offering robust designs. Optimizing with
\(\mathbf W_\nu\) adapts the finite spectral degrees of freedom to the
harmonics that remain effective under phase averaging.

\section{Conclusion}
\label{sec:conclusion}

We formulated binary QCS as direct response optimization. For a \(Q\)-query
GQSP sensor, Bayes-risk minimization over bounded degree-\(Q\) trigonometric
responses admitted an exact SDP, avoiding nonconvex control-angle optimization
while preserving realizability. Spectral factorization and GQSP synthesis
recovered the controls. Numerical results confirmed globally optimized
finite-depth responses, reliable synthesis, and robustness gains from
noise-aware redesign.

\bibliographystyle{IEEEbib}
\bibliography{strings,refs}

\end{document}